\documentclass[aps,prc,superscriptaddress,twocolumn]{revtex4-1}
\usepackage{url}
\usepackage{pstricks}
\usepackage{psfrag}
\usepackage[dvips]{graphicx}
\usepackage{xspace}
\usepackage{multirow}
\usepackage{dcolumn} %centers to the decimal point in tables using ``d''
\usepackage{amssymb}
\usepackage{bm}
\usepackage{array}
\usepackage{scrextend}
\usepackage[T1]{fontenc}
\newcommand{\mevu}{MeV/nucleon}
\newcommand{\ts}{\textsuperscript}
\newcommand{\bs}{\boldsymbol}
\newcommand{\gammaray}{\ensuremath{\gamma}-ray\xspace}
\newcommand{\gammarays}{\ensuremath{\gamma}-rays\xspace}

\newcommand{\etwop}{\ensuremath{E(2^+_1)}\xspace}
\newcommand{\efourp}{\ensuremath{E(4^+_1)}\xspace}

\newcommand{\etwopt}{\ensuremath{2^+_1 \rightarrow 0^{+}_{\mathrm{gs}}}\xspace}

\newcommand{\stwop}{\ensuremath{2^+_1}\xspace}
\newcommand{\sfourp}{\ensuremath{4^+_1}\xspace}

\newcommand{\sthreem}{\ensuremath{3^-}\xspace}

\newcommand{\ppp}{\ts{50}Ar\ensuremath{(p,p')}\ts{50}Ar}
\newcommand{\ppn}{\ts{51}Ar\ensuremath{(p,pn)}\ts{50}Ar}
\newcommand{\ptp}{\ts{51}K\ensuremath{(p,2p)}\ts{50}Ar}
\newcommand{\dali}{DALI2\ts{+}}
\newcommand{\mgamma}{\ensuremath{M_{\gamma}}}

\newcolumntype{C}[1]{>{\centering\arraybackslash}p{#1}}

\begin{document}

%\preprint{}

\title{$\boldsymbol{N=32}$ shell closure below calcium: Low-lying structure of \ts{50}Ar}
%\title{Study of the $\boldsymbol{N=32}$ shell closure below calcium: Low-lying structure of \ts{50}Ar}



%I NEED TO DOUBLE CHECK THE AUTHOR LIST-----------------------------------------------------------------------------------------------------------

\newcommand{\aatomki}{     \affiliation{Institute for Nuclear Research (Atomki), P.O. Box 51, Debrecen H-4001, Hungary}}
\newcommand{\abeijing}{    \affiliation{State Key Laboratory of Nuclear Physics and Technology, Peking University, Beijing 100871, P.R. China}}
\newcommand{\acaen}{       \affiliation{LPC Caen, ENSICAEN, Universit\'e de Caen, CNRS/IN2P3, F-14050 Caen, France}}
\newcommand{\acea}{        \affiliation{IRFU, CEA, Universit\'e Paris-Saclay, F-91191 Gif-sur-Yvette, France}}
\newcommand{\acns}{        \affiliation{Center for Nuclear Study, The University of Tokyo, RIKEN campus, Wako, Saitama 351-0198, Japan}}
\newcommand{\aewha}{       \affiliation{Department of Science Education and Department of Physics, Ewha Womans University, Seoul 03760, Korea}}
\newcommand{\agsi}{        \affiliation{GSI Helmholtzzentrum f\"ur Schwerionenforschung GmbH, Planckstr. 1, 64291 Darmstadt, Germany}}
\newcommand{\ahku}{        \affiliation{Department of Physics, The University of Hong Kong, Pokfulam, Hong Kong}}
\newcommand{\ainst}{       \affiliation{Institute for Nuclear Science \& Technology, VINATOM, P.O. Box 5T-160, Nghia Do, Hanoi, Vietnam}}
\newcommand{\aipno}{       \affiliation{IPN Orsay, CNRS and Universit\'e Paris-Saclay, F-91406 Orsay Cedex, France}}
\newcommand{\ajaea}{       \affiliation{Advanced Science Research Center, Japan Atomic Energy Agency, Tokai, Ibaraki 319-1195, Japan}}
\newcommand{\akoeln}{      \affiliation{Institut f\"ur Kernphysik, Universit\"at zu K\"oln, D-50937 Cologne, Germany}}
\newcommand{\akth}{        \affiliation{Department of Physics, Royal Institute of Technology, SE-10691 Stockholm, Sweden}}
\newcommand{\alanzhou}{    \affiliation{Institute of Modern Physics, Chinese Academy of Sciences, Lanzhou, China}}
\newcommand{\amadrid}{     \affiliation{Instituto de Estructura de la Materia, CSIC, E-28006 Madrid, Spain}}
\newcommand{\aorsay}{      \affiliation{CSNSM, CNRS/IN2P3, Universit\'e Paris-Sud, F-91405 Orsay Campus, France}}
\newcommand{\aosaka}{       \affiliation{Department of Physics, Osaka City University, Osaka 558-8585, Japan}}
\newcommand{\aoslo}{       \affiliation{Department of Physics, University of Oslo, N-0316 Oslo, Norway}}
\newcommand{\arcnp}{       \affiliation{Research Center for Nuclear Physics (RCNP), Osaka University, Ibaraki 567-0047, Japan}}
\newcommand{\ariken}{      \affiliation{RIKEN Nishina Center, 2-1 Hirosawa, Wako, Saitama 351-0198, Japan}}
\newcommand{\arikkyo}{     \affiliation{Department of Physics, Rikkyo University, 3-34-1 Nishi-Ikebukuro, Toshima, Tokyo 172-8501, Japan}}
\newcommand{\atriumph}{     \affiliation{TRIUMF, 4004 Wesbrook Mall, Vancouver BC V6T 2A3, Canada}}
\newcommand{\atitech}{     \affiliation{Department of Physics, Tokyo Institute of Technology, 2-12-1 O-Okayama, Meguro, Tokyo, 152-8551, Japan}}
\newcommand{\atohoku}{     \affiliation{Department of Physics, Tohoku University, Sendai 980-8578, Japan}}
\newcommand{\atudarmstadt}{\affiliation{Institut f\"ur Kernphysik, Technische Universit\"at Darmstadt, 64289 Darmstadt, Germany}}
\newcommand{\aunal}{       \affiliation{Universidad Nacional de Colombia, Sede Bogota, Facultad de Ciencias,\\ Departamento de F\'{\i}sica, Bogot\'a, Colombia}}
\newcommand{\aut}{         \affiliation{Department of Physics, University of Tokyo, 7-3-1 Hongo, Bunkyo, Tokyo 113-0033, Japan}}
\newcommand{\azagreb}{     \affiliation{Ru{\dj}er Bo\v{s}kovi\'{c} Institute, Bijeni\v{c}ka cesta 54, 10000 Zagreb, Croatia}}
\newcommand{\apadova}{     \affiliation{Dipartimento di Fisica e Astronomia, Universit\`a di Padova and INFN, Sezione di Padova, Via F. Marzolo 8, I-35131 Padova, Italy}}
\newcommand{\aiphc}{       \affiliation{IPHC, CNRS/IN2P3, Universit\'e de Strasbourg, F-67037 Strasbourg, France}}
\newcommand{\aumadrid}{    \affiliation{Departamento de F\'{\i}sica Te\'orica and IFT-UAM/CSIC, Universidad Aut\'onoma de Madrid, E-2804 Madrid, Spain}}
\newcommand{\aemmi}{       \affiliation{ExtreMe Matter Institute EMMI, GSI Helmholtzzentrum f\"ur Schwerionenforschung GmbH, 64291 Darmstadt, Germany}}
\newcommand{\amaxplankh}{  \affiliation{Max-Planck-Institut f\"ur Kernphysik, Saupfercheckweg 1, 69117 Heidelberg Germany}}
\newcommand{\awashington}{ \affiliation{Department of Physics, University of Washington, Seattle WA, USA}}
\newcommand{\aprisma}{     \affiliation{Institut f\"ur Kernphysik and PRISMA Cluster of Excellence, Johannes Gutenberg-Universit\"at, Mainz 55099, Germany}}
\newcommand{\alnl}{        \affiliation{Istituto Nazionale di Fisica Nucleare, Laboratori Nazionali di Legnaro, I-35020 Legnaro, Italy}}
\newcommand{\ajaveriana}{  \affiliation{Pontificia Universidad Javeriana, Facultad de Ciencias, Departamento de F\'{\i}sica, Bogot\'a, Colombia}}
\newcommand{\abarcelona}{  \affiliation{Departament de F\'{\i}sica Qu\`antica i Astrof\'{\i}sica, Universitat de Barcelona, 08028 Barcelona, Spain}}
\newcommand{\amcgill}{     \affiliation{Department of Physics, McGill University, 3600 Rue University, Montr\'eal, QC H3A 2T8, Canada}}

%first two authors fixed by data distribution
 
\author{M.~L.~Cort\'es}\ariken \alnl
%\email[email: ]{liliana.cortes@lnl.infn.it} 

%then in alphabetical order
\author{W.~Rodriguez}\aunal\ariken\ajaveriana
\author{P.~Doornenbal} \ariken
\author{A.~Obertelli} \acea \atudarmstadt
\author{J.~D.~Holt} \atriumph \amcgill
\author{J.~Men\'endez}\acns\abarcelona
\author{K.~Ogata}\arcnp \aosaka
\author{A.~Schwenk}\atudarmstadt \aemmi \amaxplankh 
\author{N.~Shimizu}\acns
\author{J.~Simonis}\aprisma
%%%\author{S. R. Stroberg}\atriumph\awashington
\author{Y.~Utsuno}\ajaea \acns
\author{K.~Yoshida}\ajaea 

%%
%the rest of the people
%%%spokepersons
%\author{P.~Doornenbal} \ariken
%\author{A.~Obertelli} \atudarmstadt \acea \ariken
%%%%local, MINOS, SAMURAI people, others

\author{L.~Achouri}\acaen
\author{H.~Baba} \ariken
\author{F.~Browne}\ariken
\author{D.~Calvet} \acea
\author{F.~Ch\^ateau} \acea
\author{S.~Chen} \abeijing \ariken
\author{N.~Chiga}\ariken 
\author{A.~Corsi} \acea %%%%%%
%\author{M.~L.~Cort\'es} \ariken
\author{A.~Delbart} \acea
\author{J-M.~Gheller} \acea
\author{A.~Giganon}\acea %%%%%%%%%%% not on shifts
\author{A.~Gillibert} \acea
\author{C.~Hilaire} \acea
\author{T.~Isobe} \ariken 
\author{T.~Kobayashi}\atohoku
\author{Y.~Kubota} \ariken \acns
\author{V.~Lapoux} \acea
\author{H.~N.~Liu}\acea\atudarmstadt\akth
\author{T.~Motobayashi} \ariken %%% Motobayashi is co-author for experiments until 2017
\author{I.~Murray} \aipno \ariken
\author{H.~Otsu} \ariken
\author{V.~Panin}\ariken
\author{N.~Paul} \acea
%\author{W.~Rodriguez}\aunal \ariken
\author{H.~Sakurai} \ariken \aut
\author{M.~Sasano} \ariken
\author{D.~Steppenbeck}\ariken
\author{L.~Stuhl}\acns\aatomki
\author{Y.~L.~Sun}\acea\atudarmstadt
\author{Y.~Togano}\arikkyo  
\author{T.~Uesaka} \ariken
\author{K.~Wimmer}\aut
\author{K.~Yoneda} \ariken %%%% Ask whether he participated
\author{O.~Aktas}\akth
\author{T.~Aumann}\atudarmstadt \agsi  % Neuland
\author{L.~X.~Chung} \ainst 
\author{F.~Flavigny}\aipno
\author{S.~Franchoo}\aipno
\author{I.~Ga\v{s}pari\'{c}}\azagreb \ariken   % Neuland
\author{R.-B.~Gerst}\akoeln
\author{J.~Gibelin}\acaen
\author{K.~I.~Hahn} \aewha
\author{D.~Kim} \aewha
\author{T.~Koiwai} \aut  %%%% Took only shifts and is master student
\author{Y.~Kondo}\atitech   % Nebula
\author{P.~Koseoglou}\atudarmstadt \agsi
\author{J.~Lee} \ahku
\author{C.~Lehr}\atudarmstadt % neuland
\author{B.~D.~Linh}\ainst 
\author{T.~Lokotko}\ahku
\author{M.~MacCormick}\aipno
\author{K.~Moschner}\akoeln
\author{T.~Nakamura}\atitech  % Nebula/SAMURAI Steering Board chair
\author{S.~Y.~Park} \aewha
\author{D.~Rossi}\atudarmstadt   % Neuland
\author{E.~Sahin} \aoslo
\author{P.-A.~S\"oderstr\"om} \atudarmstadt
\author{D.~Sohler}\aatomki  
\author{S.~Takeuchi} \atitech %% -> Yes
\author{H.~Toernqvist}\atudarmstadt \agsi % Neuland
\author{V.~Vaquero}\amadrid 
\author{V.~Wagner}\atudarmstadt % Neuland
\author{S.~Wang}\alanzhou
\author{V.~Werner}\atudarmstadt
\author{X.~Xu} \ahku
\author{H.~Yamada}\atitech % Nebula
\author{D.~Yan} \alanzhou
\author{Z.~Yang} \ariken 
\author{M.~Yasuda}\atitech  % Nebula
\author{L.~Zanetti}\atudarmstadt % Neuland

%\thanks{}


\date{\today}

\begin{abstract}
  Low-lying excited states in the $N=32$ isotope \ts{50}Ar were investigated by in-beam $\gamma$-ray spectroscopy following  proton- and neutron-knockout, multi-nucleon removal, and  proton inelastic scattering at the RIKEN Radioactive Isotope Beam Factory.
  The energies of the two  previously reported transitions have been confirmed, and five additional states are presented for the first time, including a candidate for a \sthreem  state.
  The level scheme built using $\gamma\gamma$ coincidences was compared to shell-model calculations in the $sd-pf$ model space, and to ab initio predictions based on chiral two- and three-nucleon interactions.
  Theoretical proton- and neutron-knockout  cross sections  suggest that two of the new transitions correspond to $2^+$ states, while  the previously proposed \sfourp state could also correspond to a $2^+$ state.
%  The agreement of the experimental data with the theoretical predictions strengthens our understanding of the shell evolution in this region.

 \end{abstract}

%\pacs{}
%\keywords{}

\maketitle


\section{Introduction}

Our understanding of the  atomic nucleus has as one of its  cornerstones the concept of shell structure, in which the location of single-particle orbitals defines shell closures and  associated  magic numbers.
Experimental evidence collected in the past decades, particularly since the advent of radioactive ion beams, has shown that shell structure undergoes significant changes for isotopes far from stability~\cite{Sorlin_PPNP61_2008}.
Examples of this shell-evolution are the onset  of $N=16$ as a magic number for O isotopes~\cite{Ozawa_PRL_84_2000,Hoffman_PLB_672_2009,Kanungo_PRL102_2009} and the disappearance of the canonical magic number $N=20$ around \ts{32}Mg~\cite{Detraz_PRC_19_1979,Motobayashi_PLB_346_1995}.

A particularly interesting case to study shell evolution is the region around the Ca isotopes between $N=28$ and $N=40$, where the development of shell closures for $N=32$ and $N=34$ has  recently gained significant attention. % from both the experimental and theoretical points of view. 
In the Ca isotopes, the $N=32$ sub-shell closure was first evidenced by its relatively high \etwop energy~\cite{Huck_PRC_31_1985}, and confirmed by two-proton knockout cross section~\cite{Gade_PRC_74_2006} and mass measurements~\cite{Wienholtz_Nature498_2013}.
In turn, the first suggestion of the  $N=34$ shell closure on \ts{54}Ca was also provided by the \etwop measurement~\cite{Steppenbeck_Nature502_2013} and confirmed by systematic mass measurements~\cite{Michimasa_PRL_121_2018} and neutron-knockout reactions~\cite{Chen_PRL_123_2019}.

The persistence of these shell closures below and above $Z = 20$ has also been widely investigated. 
The preservation of the   $N=32$ shell closure above Ca has been determined in Ti and Cr via spectroscopy~\cite{Janssens_PLB_546_2002,Prisciandaro_PLB_510_2001}, reduced transition probabilities~\cite{Dinca_PRC_71_2005,Burger_PLB_622_2005}, and  precision mass measurements~\cite{Leistenschneider_PRL_120_2018}.
On the other hand, the $N = 34$ shell closure has been suggested to disappear above Ca~\cite{Liddick_PRL_92_2004}.
This is in contrast with the recently reported first spectroscopy  measurement on \ts{52}Ar, where  the experimental value of   \etwop  suggests the conservation of the $N=34$ shell closure for $Z=18$~\cite{Liu_PRL_122_2019}.

The first spectroscopy of \ts{50}Ar showed a relatively high  \etwop energy of 1178(18)~keV~\cite{Steppenbeck_PRL_114_2015}. 
In that study, apart from the \etwop,  a  \efourp was tentatively assigned, although the limited statistics prevented a firmer conclusion~\cite{Steppenbeck_PRL_114_2015}. No further spectroscopic information is available for this very exotic nucleus. 
% suggesting the  conservation of the magic character of $N=32$ below Ca~\cite{Steppenbeck_PRL_114_2015}. 
The increase of the \etwop with respect to neighboring  isotopes has been interpreted as an indication of a sizable $N=32$  gap along the Ar isotopic chain, therefore maintaining this sub-shell closure below \ts{52}Ca~\cite{Steppenbeck_PRL_114_2015}.

%Based in the agreement of the results with shell-model calculations, the  conservation of the magic character of $N=32$ below Ca was suggested~\cite{Steppenbeck_PRL_114_2015}. 
%%---

From a theoretical point of view, the tensor-force-driven shell evolution has been used to explain the appearance of the $N=32$ and $N=34$  shell closures~\cite{Otsuka_PRL_95_2005}. In this framework, the reduction of the attractive proton-neutron interaction between the $\pi f_{7/2}$ and the $\nu f_{5/2}$ single-particle orbitals results in a separation between these levels  and the  formation of substantial neutron gaps. % at $N=32$ and $N=34$.
Calculations  including this effect~\cite{Otsuka_PRL_104_2010,Utsuno_JPSCP_6_2014}  successfully reproduce the \etwop of Ar isotopes~\cite{Steppenbeck_PRL_114_2015,Liu_PRL_122_2019} and suggest the magnitude of the $N = 34$ sub-shell closure in \ts{52}Ar to be around 3~MeV. 

The significance  of  three-nucleon forces (3NFs) in the description of neutron-rich isotopes has also been studied~\cite{Otsuka_FewBodySys_54_2013,Holt_JOPG_40_2013}, and the relevance of this contribution to obtain an accurate description of the spectroscopic properties of Ca isotopes has been highlighted~\cite{Holt_PRC_90_2014}.
In particular, ab initio calculations with the valence-space in-medium similarity renormalization group  (VS-IMSRG) method~\cite{Tsukiyama_PRC_85_2012,Hergert_PhysRep_621_2016,Stroberg_AnnRevNuclPartSci_69_2019} including 3NFs have provided a satisfactory description of the \etwop along the Ar isotopic chain~\cite{Liu_PRL_122_2019}.

Our understanding of the nature of these  sub-shell closures relies on the interpretation provided by the theoretical calculations.  The validity of this  picture can be further tested by studying its agreement with other nuclear properties, for example, the energies of low-lying states beyond the \stwop.  To get a better insight into the structure at the $N=32$ shell closure below Ca, the present work reports low-lying states in \ts{50}Ar populated  following direct and indirect reactions.

\section{Experiment}

The experiment was performed  at the Radioactive Isotope Beam Factory, operated by the RIKEN Nishina Center and the Center for Nuclear Study  of the University of Tokyo. 
A  \ts{70}Zn beam with an energy of  345~\mevu~   and an average intensity of  240~pnA was fragmented on a   3-mm thick~Be target  to produce the secondary beam cocktail. 
Fragments of interest were selected by the BigRIPS separator~\cite{Kubo_PTEP2012_2012} using the $B\rho - \Delta E - B\rho$ technique. Event-by-event identification was obtained by an energy-loss measurement  in an ionization chamber,  position and angle measurements  with parallel plate avalanche counters, 
 at different focal planes, 
and the time-of-flight measured between two plastic scintillators~\cite{Kubo_PTEP2012_2012}.
The selected isotopes were  focused in front of the SAMURAI dipole magnet~\cite{Kobayashi_NIM_317_2013}, where  the 151.3(13)-mm long liquid hydrogen target of MINOS~\cite{Obertelli_EPJA50_2014, Santamaria_NIMA_905_2018}  was placed. Thanks to the use of a time projection chamber  surrounding the target, it was possible to reconstruct the reaction vertex with a resolution of 2~mm~($\sigma$)~\cite{Santamaria_NIMA_905_2018}.
Following the  reactions in the target,  ions were identified using the SAMURAI magnet and associated detectors. Positions and angles were measured  at two multi-wire drift chambers placed in front  and behind  the magnet, the time-of-flight was obtained from a scintillator placed in front of the target and a  hodoscope located downstream of SAMURAI, which also provided  an energy loss measurement from which the atomic number was inferred~\cite{Kobayashi_NIM_317_2013}.
%Two large-acceptance plastic scintillator arrays, the NeuLAND demonstrator~\cite{Aumann_NeuLand} and NEBULA~\cite{Kobayashi_NIM_317_2013,Nakamura_NIMB_376_2016}, were placed at zero degree, about 12~m behind the target, to measure the neutrons emitted in the reaction. For the present study the neutron information was not employed. 

The high-efficiency $\gamma$-ray detector  array DALI2\ts{+}~\cite{Takeuchi_NIMA763_2014,Murray_RAPR_2018},  composed of 226 NaI(Tl) detectors, was placed around MINOS to detect de-excitation $\gamma$ rays.  The array, which covered  detection angles between  $\sim$12\ts{$\circ$} and $\sim$118$^{\circ}$ with respect to the center of the target,  was calibrated in energy  using standard  \ts{60}Co, \ts{88}Y, \ts{133}Ba, and \ts{137}Cs sources.
The full-energy-peak efficiency of the array, determined using a detailed GEANT4~\cite{Agostinelli_NIMA506_2003} simulation,  was 30\% at 1~MeV with an energy resolution of 11\% for  a   source moving at a velocity of  0.6$c$.
Previous results and further details  from the same experiment can be found in Refs.~\cite{Liu_PRL_122_2019, Chen_PRL_123_2019, Cortes_PLB_800_2019, Sun_PLB_802_2020}.

\section{Results}

\setlength\extrarowheight{2.5pt}
\begin{table}[b]
  {%\footnotesize
    \caption{Inclusive cross sections ($\sigma_{\text{inc}}$) obtained for each of the reaction channels populating  \ts{50}Ar. The total number of events measured in each channel, the mean incident beam energy ($E_{beam}$), as well as the efficiency of MINOS ($\varepsilon_{\text{\tiny{MINOS}}}$) are  listed.\label{tab:reactions}}
    \begin{ruledtabular}
 % \begin{tabular}{|cccc|}
      \begin{tabular}{ccccc}
%  \begin{tabular}{|C{3cm}C{1cm}C{2cm}C{1.5cm}|}
   \multirow{2}{*}{Reaction} & \multirow{2}{*}{Events} & $E_{beam}$       & $\varepsilon_{\text{\tiny{MINOS}}}$   &$\sigma_{\text{inc}}$ \\
             &        & \small{(MeV/$u$)}   &                            (\%)  & (mb)\\\hline
\ts{52}Ca$(p,3p)$\ts{50}Ar  &   132  & 266   &  99(12)   &  0.09(1)\\ 
\ts{53}Ca$(p,3pn)$\ts{50}Ar &   999  & 258   &  82(8)    &  0.33(3)\\ 
\ts{54}Ca$(p,3p2n)$\ts{50}Ar&   1393 & 251   &  88(8)    &  0.81(7)\\ 
\ts{55}Ca$(p,3p3n)$\ts{50}Ar&   790  & 247   &  85(3)    &  1.04(4)\\ \hline
\ts{51}K$(p,2p)$\ts{50}Ar   &  28177 & 257   &  92(2)    &  3.9(1)\\  %*
\ts{52}K$(p,2pn)$\ts{50}Ar  &  13900 & 250   &  91(3)    &  8.7(3)\\
\ts{53}K$(p,2p2n)$\ts{50}Ar &  5837  & 245   &  86(6)    &  12.2 (8)\\\hline
\ts{51}Ar$(p,pn)$\ts{50}Ar  &  1214  & 241   &  70(2)    & 45(2)\\ 
%\ts{50}Ar$(p,p)$\ts{50}Ar        &  743703 & 16(4)    & 26(1)\footnote{Corresponding to elastic and inelastic scattering}\\ 
   \end{tabular}
\end{ruledtabular}
  } 
\end{table}

Low-lying states in \ts{50}Ar were populated by direct and indirect reactions. 
For each reaction channel inclusive cross sections  were obtained using the  effective transmission of \ts{50}Ar (which includes the efficiency of the  beam line detectors and the beam losses in the detectors and the target) measured to be  56.7(15)\%, and the efficiency of MINOS for each reaction.
%No condition on the neutrons being emitted  was considered at this point, therefore the reported  cross sections also include  the contribution from  possible neutron evaporation following the direct reaction.
Table \ref{tab:reactions} summarizes the number of events  in each reaction channel, the mean incident beam energy, the experimental efficiency of MINOS, and the corresponding  inclusive cross sections. 

Doppler corrected $\gamma$-ray spectra were obtained using the reaction vertex and the velocity of the fragment reconstructed with  MINOS.
Peak-to-total ratio and  detection efficiency  were improved by adding-up the energies of $\gamma$ rays deposited in  detectors up to 10~cm apart. To reduce the contribution of the  low-energy atomic background, $\gamma$ rays with energies below 100~keV in the laboratory frame of reference were not considered for the add-back.

\begin{figure}[b!]
 \includegraphics[width=0.46\textwidth]{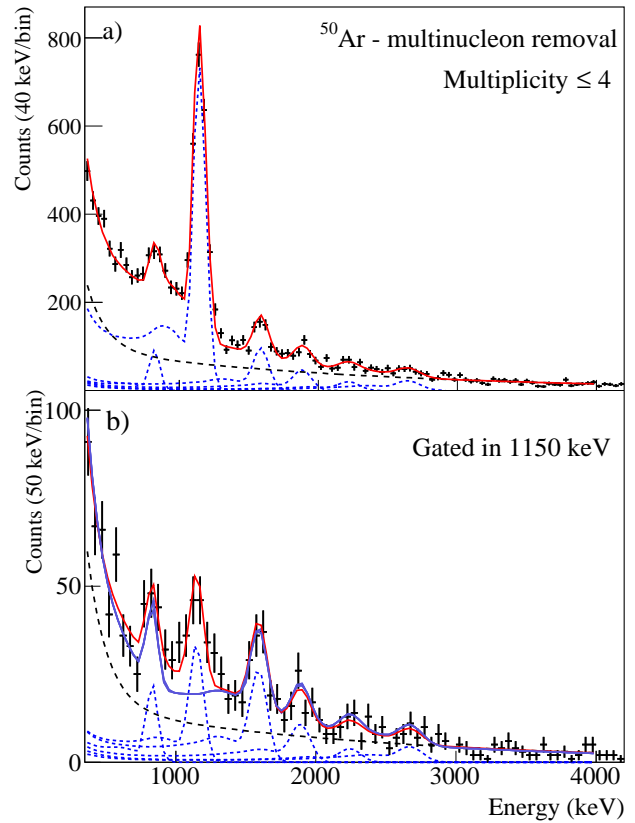}%
 \caption{a) Doppler-corrected \gammaray spectrum obtained for \ts{50}Ar populated from multinucleon removal reactions. The dashed blue  lines represent the simulated responses of DALI2\ts{+} to the  different transitions, the dashed black line shows the fitted double-exponential background, and the solid red line shows the total fit. b)~ \gammaray spectrum gated at $\sim1150$~keV. The best fit is shown by the solid red line while the expected counts are shown by the darker line.\label{fig:spectrum_mult}}
\end{figure}

Figure~\ref{fig:spectrum_mult}a) shows the Doppler-corrected spectrum obtained following multinucleon removal reactions, when the  \gammaray multiplicity (\mgamma)  was limited to  a maximum of four.
The spectrum was fitted with  simulated response functions of the \dali array and  a double exponential function used to model the low- and high-energy background. The slopes of the two  exponential functions were fixed by independent fits of  the high- and low-energy regions.
Six transitions at 826(7)(8)~keV, 1151(1)(12)~keV, 1593(6)(16)~keV, 1892(11)(19)~keV, 2227(19)(22)~keV, and 2657(21)(27)~keV provided the best fit to the spectrum. 
The first reported uncertainty corresponds to the statistical error from the fit, while the second is the systematic error arising from the calibration of the \gammaray detectors  and the possible lifetime of the states.  %Regarding this latter error,  following Ref.~\cite{Raman_ADNT_78_2001} a lifetime below 6~ps was considered.
To place the observed transitions in a level scheme,  $\gamma\gamma$ coincidences were investigated.
Figure~\ref{fig:spectrum_mult}b) displays the   \gammaray spectrum gated  between 1090~keV and 1210~keV. A single background gate between 3000~keV and 4000~keV was used. Due to the many transitions observed in the spectrum it was not possible to place a more appropriate background gate.  As a results, the   transition where the gate was placed could not be completely removed by the background subtraction. Hence, the possibility of a doublet cannot be fully excluded.
The best fit to the resulting spectrum, shown by the red line, was obtained by using the same response functions as in Fig~\ref{fig:spectrum_mult}a), suggesting that all the transitions are coincident with the one at $\sim$1150~keV.
Calculations on the expected number of counts in the coincidence spectrum obtained based on the area of the gate and the efficiency of \dali, are consistent with the observations, as shown by the blue line Fig.~\ref{fig:spectrum_mult}b).  %In this curve,  the expected peak amplitudes were fixed as well as the slopes of the exponential background functions and the only free parameters for the fit were the amplitudes of these  the exponential functions.

\begin{figure}[t]
 \includegraphics[width=0.46\textwidth]{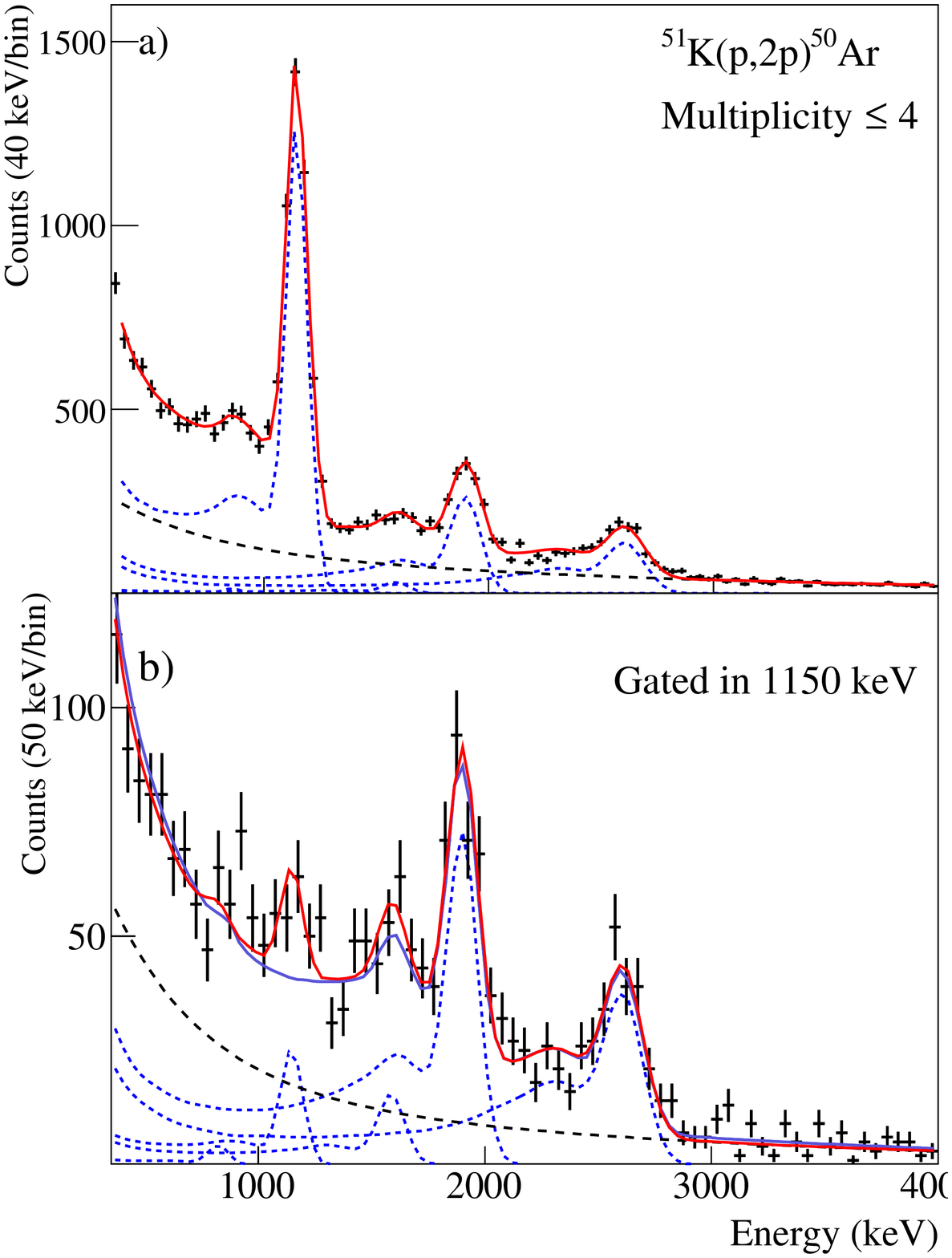}%
 \caption{Same as Fig.~\ref{fig:spectrum_mult}, but for the \ptp~reaction. \label{fig:spectrum_p2p}}
\end{figure}

Figure~\ref{fig:spectrum_p2p}a) shows the  Doppler-corrected spectrum obtained for \ts{50}Ar produced by the proton-knockout reaction. % with $\mgamma\leq4$. 
A total of four peaks provided the best fit to the spectrum. 
The transition energies deduced from this spectrum are  1150(1)(11)~keV, 1592(23)(16)~keV, 1905(3)(19)~keV, and 2618(6)(26)~keV.
The spectrum resembles the one observed for the multinucleon removal, and in fact all of the transitions observed seem to correspond within uncertainties to transitions also present in  Fig.~\ref{fig:spectrum_mult}. In this case, however, the intensity of the transition around  $\sim1600$~keV is smaller, while the transitions at $\sim1900$~keV and $\sim2600$~keV are more intense. The transition observed in Fig.~\ref{fig:spectrum_mult} at $\sim824$ was observed with a significance below $1\sigma$, and the one at  $\sim2230$~keV was not visible in this spectrum. 
The projection of the $\gamma\gamma$ matrix gated around $\sim1150$~keV is shown in Fig.~\ref{fig:spectrum_p2p}b).
The best fit to the spectrum was obtained using the same  four response functions used to fit the total spectrum, indicating that the  transitions are coincident with the one at $\sim 1150$~keV.  As in the case of Fig.~\ref{fig:spectrum_mult}a), it was not possible to completely remove the transition where the gate was placed by background subtraction. 
%This is consistent with the observation in Fig.~\ref{fig:spectrum_mult} and, thanks to the higher statistics, strengthens it. 
It is noted that for the \ptp~ and the multinucleon removal reactions, gates around  $\sim1600$~keV, $\sim1900$~keV, and $\sim2600$~keV only showed the reciprocal coincidence of these  transitions with the one at  $\sim1150$~keV.

\begin{figure}[t]
 \includegraphics[width=0.46\textwidth]{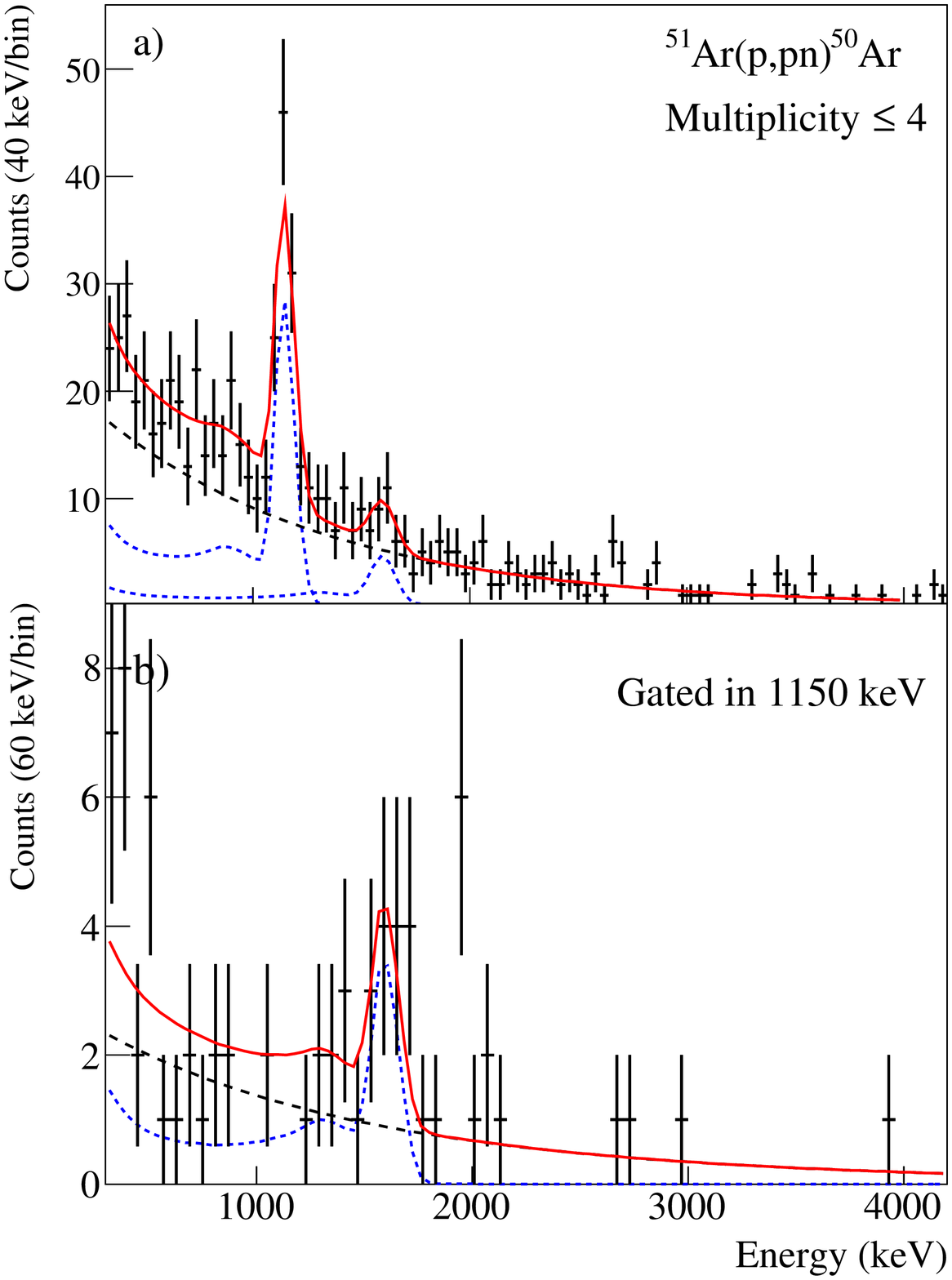}%
 \caption{Same as Fig.~\ref{fig:spectrum_mult} for  the \ppn~ reaction.  \label{fig:spectrum_ppn}}
\end{figure}

\begin{figure}[t]
 \includegraphics[width=0.46\textwidth]{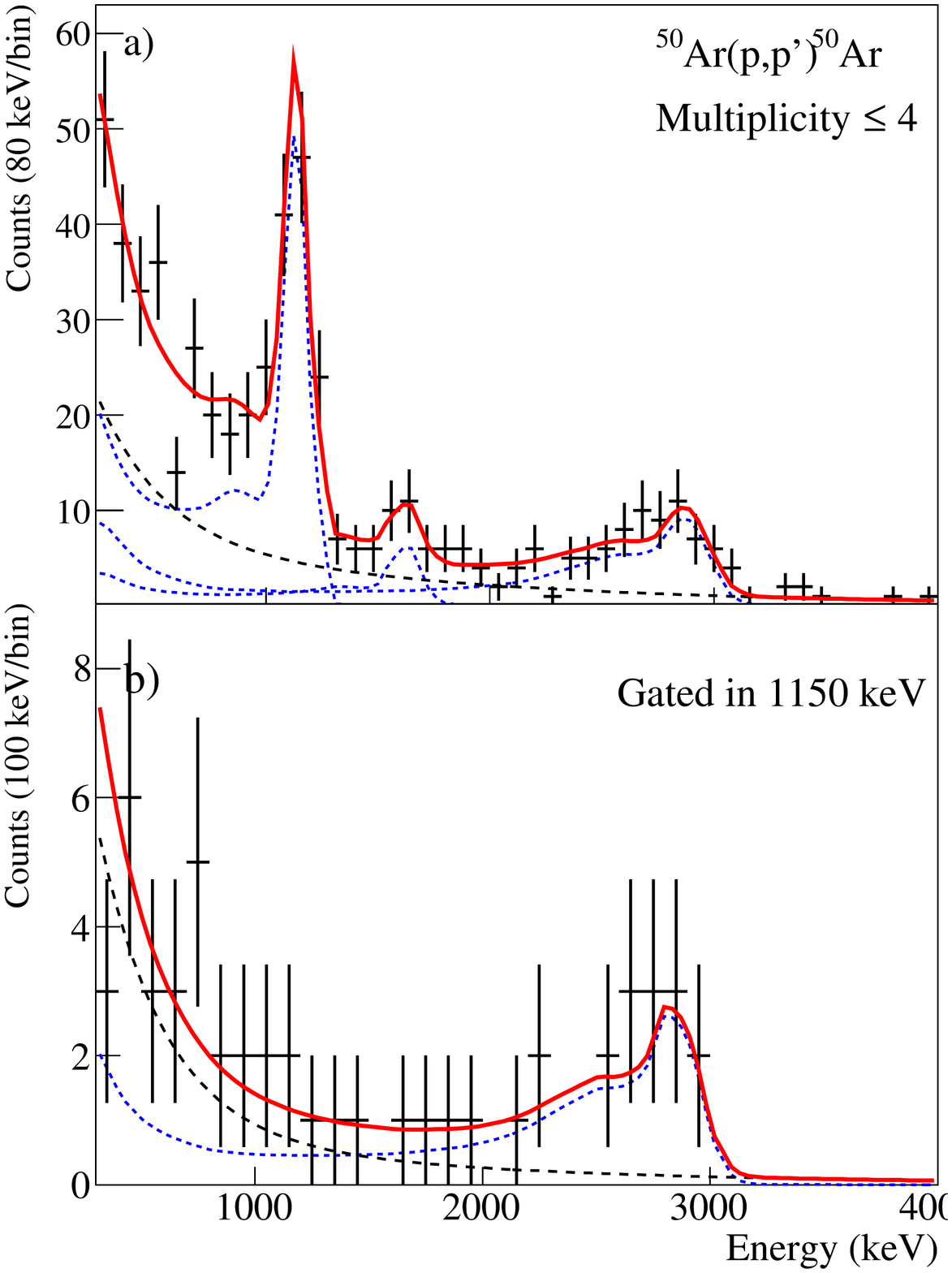}%
 \caption{Same as Fig.~\ref{fig:spectrum_mult}  for the \ppp~ reaction. \label{fig:spectrum_ppp}}
\end{figure}

The Doppler-corrected spectrum corresponding to the  \ppn~ reaction  is displayed in Fig.~\ref{fig:spectrum_ppn}a).
Two  peaks are visible in the spectrum with a significance above $2\sigma$. The best fit yields  transition energies of 1150(8)(11)~keV, and 1602(31)(16)~keV. 
The $\gamma\gamma$ analysis shown in Fig.~\ref{fig:spectrum_ppn}b) clearly establishes the existence of the peak at $\sim1600$~keV, and  shows that it is coincident with the one at $\sim1150$~keV.  
The energies observed in these spectrum are consistent with the ones obtained previously, suggesting the population of the same levels. %, although due to the limited resolution,  it cannot be ruled out that they correspond to different transitions. 
%Although in Fig.~\ref{fig:spectrum_ppn}a) an excess of counts is observed  around 2~MeV, the fit of a second transition does not improve the $\chi^2$ of the fit. 
%Furthermore, based on the efficiency of DALI2$^+$, a peak with this intensity in the coincidence spectrum should also be visible in  singles, which is not the case. 

Figure~\ref{fig:spectrum_ppp}a) shows the Doppler-corrected spectrum obtained for the \ppp ~reaction. Three transitions are visible and the transition energies obtained for this case are  1138(8)(11)~keV, 1626(33)(16)~keV, and 2890(31)(29)~keV.
The background-subtracted coincidence spectrum, in Fig.~\ref{fig:spectrum_ppp}b), shows that the transition at 2890~keV is coincident with the one at 1138~keV. No coincidence between the transitions at 1626~keV and 1138~keV was observed, which can be attributed to the  reduced statistics. %, although it is possibile  that the peak at $\sim~1626$~keV does not corresponds to peak  previously observed at $\sim~1590$~keV. %A subsequent measurement using a high-resolution \gammaray array is necessary to determine 

\begin{figure*}[t]
 \includegraphics[width=\textwidth]{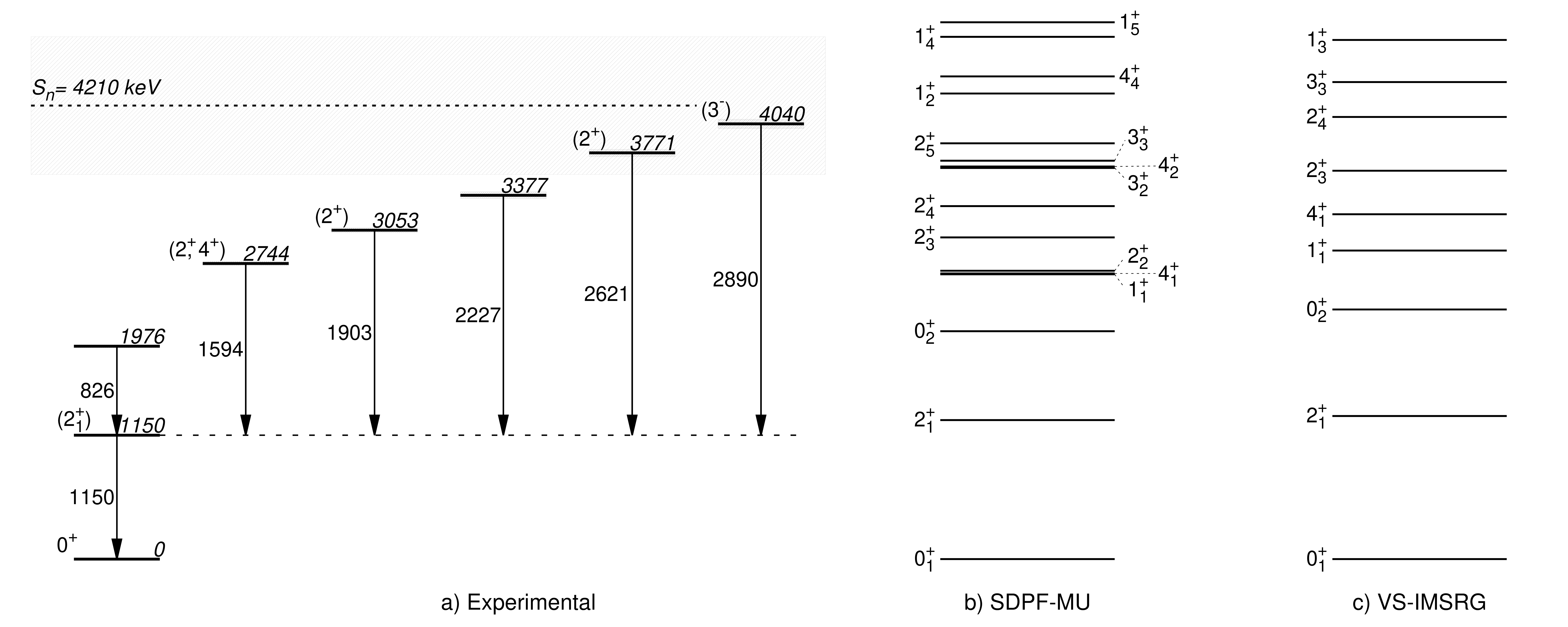}%
 \caption{a) Experimental level scheme for \ts{50}Ar deduced in the present work. Level and transition energies are given by the italic and regular fonts, respectively.  The calculated neutron separation energy, $S_n$, is indicated~\cite{Wang_CPC_41_2017}. Uncertainties in the energy levels are displayed as shaded areas. Parts b) and c) display  predictions for low-lying states in \ts{50}Ar by the SDPF-MU shell model and  VS-IMSRG calculations, respectively. 
   \label{fig:levels}}
\end{figure*}

%\section{Discussion}

Based on the $\gamma\gamma$ analysis discussed above, the tentative level scheme shown in Fig.~\ref{fig:levels}a) was constructed. 
The energies of low-lying states in \ts{50}Ar were calculated as the weighted average of the values obtained from the different reactions, when applicable.
The weights were determined based only on the statistical uncertainty, and the systematic error was added in quadrature.
Being the one with the highest intensity, the 1150(12)~keV transition was placed decaying directly into  the ground state. This transition  agrees, within error bars, with the one at  1178(18)~keV reported in Ref.~\cite{Steppenbeck_PRL_114_2015}, where it was tentatively assigned to the \etwopt transition.
The transitions at 826(9)~keV, 1594(16)~keV, 1903(19)~keV, 2227(30)~keV, and 2621(32)~keV, were placed feeding the  (\stwop) state,  in parallel to each other, depopulating states at 1976(15)~keV, 2744(20)~keV, 3053(23)~keV,  3377(32)~keV, and 3771(34)~keV,  respectively.
The  transition at 1594(16)~keV  agrees  with the one at  1582(38)~keV reported in Ref.~\cite{Steppenbeck_PRL_114_2015}, where a \sfourp assignment was suggested. 
%It is noted that for this transition, only the muti-nucleon removal and the \ptp~ reactions were used to obtain the weighted average.
%The low statistics achieved in the previous experiment might help to explain why the measurements diverge, although the existence of more than one level at this energy cannot be ruled out. 
The transition at 2890(42)~keV observed in the \ppp~reaction was placed on top of the \stwop state, depopulating a level at 4040(44)~keV.
It has been  shown by previous measurements on \ts{46}Ar~\cite{Riley_PRC72_2005}, \ts{50}Ca~\cite{Riley_PRC90_2014}, and \ts{54}Ti~\cite{Riley_PRC_96_2017}, that proton inelastic scattering populates preferentially $2^+$, $4^+$, and \sthreem states,  therefore a \sthreem spin and parity can be  reasonably assigned to this level. The spin assignment for the  2744(20)~keV level, also observed in this reaction,  could then be either  $2^+$ or $4^+$.  Further discussion on the possible spin and parity assignments for the levels obtained in this work will be presented below.
%For the transition at 2316(35)~keV observed in Fig.~\ref{fig:spectrum_mult},  the low statistics and the  lack of more reliable information prevented placement in the level scheme. 

%feeding

\begin{table}[t]
  \caption{Energies of the  low-lying states in \ts{50}Ar measured in this work. The adopted levels were calculated as the weighted average of the results obtained for different reaction channels, when possible. Observed exclusive cross sections, $\sigma_{exp}$, for the direct reactions are reported.\label{tab:transitions}}
  \begin{ruledtabular}
    \begin{tabular}{|ccc|}
%    \begin{tabular}{|cccc|}
% Energy (keV) & $\sigma^{(p,2p)}_{exp}$ (mb) & $\sigma^{(p,pn)}_{exp}$ (mb)& $\sigma^{(p,p')}_{exp}$ (mb)\\\hline
 Energy (keV) & $\sigma^{(p,2p)}_{exp}$ (mb) & $\sigma^{(p,pn)}_{exp}$ (mb)\\\hline
 0                  &  $\leq$1.2(2)                          & $\leq$26(4)                 \\ %                  & -- \\
 1150(12)     &  0.8(2)                                    & $\leq$15(4)                       \\ %           &  1.9(5)  \\ 
 1976(15)     &  --                                             & --                           \\ %              &  -- \\
 2744(20)      &  0.10(3)                                  & $\leq$5(2)                        \\ %             &  0.7(4) \\
 3053(23)       &  1.0(1)                                  & --                          \\ %              &  -- \\
 3377(32)       & --                                          & --                       \\ %                  & --\\
 3771(34)       &  0.8(1)                                  & --                          \\ %              &  --  \\
 4040(44)       &  --                                           & --                     \\ %                   &  2.5(5)  \\
  \end{tabular}
\end{ruledtabular}
\end{table}

For the direct reactions, exclusive cross sections to populate each observed state were obtained from the fitted \gammaray intensities.
Table~\ref{tab:transitions} summarizes the adopted level  energies and  exclusive cross sections obtained  in this work.
Based on simulated angular distributions of the \gammarays, an additional uncertainty of 4\% has been included  to account for possible alignment of the states. 
%Assuming that no other state was populated, 
The ground state cross section  was calculated by subtracting the exclusive cross sections from  the  inclusive one reported in Table~\ref{tab:reactions}.
%For the calculations, the feeding was subtracted according to the proposed level scheme.
The high  background level, low statistics and limited resolution of \dali  could prevent the observation of low-intensity, high-energy  transitions feeding directly the $0^+_{\mathrm{gs}}$  state, therefore the ground-state cross section is prone to be overestimated. 
In addition, it was not possible  to disentangle between the direct \ppn~ reaction and the scattering followed by neutron emission, \ts{51}Ar$(p,p')$\ts{51}Ar$\rightarrow$ \ts{50}Ar $ +~n$, therefore, all the  cross sections for this channel  are to be considered as an upper limit.
%As in the spectrum of the  \ppp~ reaction no coincidence between the transitions at 1150~keV and 1630~keV was observed, an intermediate value between the 100\% and no feeding was assumed, and the error bar increased to cover both scenarios. 

%---------------------------

%Based on the expected neutron separation energy, $S_n=4210(640)$~keV~\cite{Wang_CPC_41_2017}, only states  up to $\sim$~4.6~MeV are displayed in Fig.~\ref{fig:levels}.

%For the case of the \ptp~reaction, the  $J^{\pi}=3/2^+$ ground state for \ts{51}K was employed~\cite{Papuga_PRL_110_2013,Papuga_PRC_90_2014} and knockout from the $sd$ shell was considered, leading to the population of positive parity states exclusively. 
%For the case of the  \ppn~ reaction  the predicted  ground-state spin  of $1/2^-$ for \ts{51}Ar  was assumed. %, leading to possible knockout from the $pf$ shell. 

\section{Discussion}

%To compare the experimental level scheme with theory, 
Predictions for the energies  of low-lying states in \ts{50}Ar were obtained within  the shell-model framework using the SDPF-MU effective interaction~\cite{Utsuno_PRC_86_2012} and considering  the full $sd$ and $pf$ model space for protons and neutrons.
%, as shown in  Fig.~\ref{fig:levels}b). 
The original Hamiltonian was modified~\cite{Utsuno_JPSCP_2014} using experimental data on exotic Ca~\cite{Steppenbeck_Nature502_2013} and K~\cite{Papuga_PRL_110_2013} isotopes. 
These calculations have previously provided  good agreement with the  experimental \etwop and \efourp  energies in neutron-rich Ar isotopes~\cite{Steppenbeck_PRL_114_2015} and  suggest a $N=32$ gap of  $\sim~3$~MeV for \ts{50}Ar. Although this gap is predicted to be of similar magnitude as for \ts{52}Ca, the wave function of the \stwop state for \ts{50}Ar turns out  to be  more mixed than the one for \ts{52}Ca, making the effect of this shell closure less evident~\cite{Steppenbeck_PRL_114_2015}.
%The predicted level scheme is displayed in Fig.~\ref{fig:levels}b).

Calculations were also performed using the ab initio VS-IMSRG approach using   %, producing the level scheme shown in Fig.~\ref{fig:levels}c).
%For these calculations, 
the chiral NN+3N interaction labeled 1.8/2.0 (EM) in Refs.~\cite{Simonis_PRC_96_2017,Hebeler_PRC_83_2011}.
This NN+3N interaction is based on chiral effective field theory~\cite{Epelbaum_RevModPhys_81_2009,Hammer_RevModPhys_85_2013}, a low-energy effective theory of quantum chromodynamics, with low-energy constants fitted to the properties of the lightest nuclei up to $^4$He.
The same chiral interaction has  been successfully used to study \etwop in the Ar isotopic chain~\cite{Liu_PRL_122_2019}, as well as excitation spectra from oxygen~\cite{Ciemala_PhysRevC_101_2020} to nickel~\cite{Taniuchi_Nature_569_2019} and tin~\cite{Morris_PRL_120_2018} isotopes.
For the model spaces, the  $sd$  space was considered for the  protons, and the $pf$  for the  neutrons, preventing the calculation of negative parity states.
As in previous works~\cite{Liu_PRL_122_2019,Ciemala_PhysRevC_101_2020,Taniuchi_Nature_569_2019,Morris_PRL_120_2018}  the IMSRG(2) approximation, where all induced operators are truncated at the two-body level, was employed.
The VS-IMSRG was used to decouple a valence-space Hamiltonian, which captures 3N forces between valence nucleons via an ensemble normal ordering, for each nucleus of interest~\cite{Stroberg_PRL_118_2017}.

Spectroscopic factors, $C^2S$,   were  calculated within each model. For the case of the \ptp~reaction, the  $J^{\pi}=3/2^+$ ground state for \ts{51}K was employed~\cite{Papuga_PRL_110_2013,Papuga_PRC_90_2014} and knockout from the $sd$ shell was considered, leading to the population of positive parity states exclusively. 
For the case of the  \ppn~ reaction  the predicted  ground-state spin  of $1/2^-$ for \ts{51}Ar  was assumed. %, leading to possible knockout from the $pf$ shell. 
Figs. ~\ref{fig:levels}b) and ~\ref{fig:levels}c) show the  level scheme obtained from the  calculations where only positive-parity states with  calculated $C^2S \geq 0.1 $ for the \ptp~ or \ppn~ reactions are displayed.  
%At this point only  positive-parity states are considered. 
The predictions for the \sthreem state based on the SDPF-MU Hamiltonian  will be discussed afterwards. 
%%change to intro:
%%%The \etwop of \ts{50}Ar is accurately reproduced by the two calculations. The increase on this energy with respect to neighboring  isotopes has been interpreted as an indication of a sizable $N=32$  gap along the Ar isotopic chain, therefore maintaining this sub-shell closure below \ts{52}Ca~\cite{Steppenbeck_PRL_114_2015}.
%%%Our understanding of the nature of this sub-shell closure relies on the interpretation provided by the theoretical calculations.  The validity of this  picture can be further tested by studying its agreement with other nuclear properties, for example, the energies of low-lying states beyond the \stwop and \sfourp states. 

%As can be seen from Fig.~\ref{fig:levels}, t
The \etwop of \ts{50}Ar is accurately reproduced by both calculations, and a 
 $0^+_2$ is predicted to be the next excited state.  The experimental level at 1976(15)~keV has a good agreement with this state. 
It is noted that  the SPDF-MU calculations predict the $0^+_2$ state of \ts{56}Cr to be 1982.1~keV, in fair agreement with the tentative  experimental level at  1674.5(4)~keV~\cite{Mantica_PRC_67_2003}. %No other $0^+_2$ states have been reported for lighter $N=32$ isotones. 
The structure at higher energies also  presents many similarities: 
The  next levels predicted to be populated are the  \sfourp,   $1^+_1$, $2^+_3$  and $2^+_4$.
However, the  energies predicted in the VS-IMSRG approach are modestly higher than in the  SDPF-MU calculations.  
By enlarging the configuration space of this theoretical framework to include the $sd-pf$ orbitals for protons and neutrons, additional excited states may appear at lower energies. 
The  SDPF-MU calculations also predict significant population of more levels, in particular of   the $2^+_2$  and states with spin and parity $1^+$ and  $4^+$.

\setlength\extrarowheight{2pt}
\begin{table*}[t]
  \caption{Calculated spectroscopic factors and cross sections for the states  populated in the   \ptp ~reaction.  \label{tab:ptp}}
     
  \begin{ruledtabular}
  \begin{tabular}{|c||c|ccc|c||c|ccc|c|}
                                                            & \multicolumn{5}{c||}{\bf{SPDF-MU}}        &  \multicolumn{5}{c|}{\bf{VS-IMSRG}}  \\\hline
\multirow{2}{*}{{\bf J}\bs{$^\pi$}} & \multirow{2}{*}{{\bf E(keV)}}   &\multicolumn{3}{c|}{\bs{$C^2S~$}} & \bs{$\sigma_{\text{\bf theo}}$} &   \multirow{2}{*}{{\bf E(keV)}}  &\multicolumn{3}{c|}{\bs{$C^2S~$}} & \bs{$\sigma_{\text{\bf theo}}$} \\
                                                          &                                               	 &\bs{$1s_{1/2}$} & \bs{$0d_{3/2}$} & \bs{$0d_{5/2}$} & {\bf (mb)}	    &   & \bs{$1s_{1/2}$} & \bs{$0d_{3/2}$} & \bs{$0d_{5/2}$} & {\bf (mb)}\\\hline
$0_1^+ $   &  0          & --	   &   0.30 &  --	    & 0.46  &  0        & --     & 0.21   & --    &  0.33 \\ 
$2_1^+ $   &  1291    &0.23 &	0.38 & 0.01  & 1.00  & 1328  & 0.16 & .21 & 0.02  &  0.62\\	 
$4_1^+ $   &  2651    &--	   &   --	&  0.10  & 0.18  & 3201   &  --     &   -- & 0.15  & 0.25 \\	 
$2_3^+ $   &  2986    &0.17 &	0.07 &   --     & 0.39  &             &           &   &          &  \\
$2_4^+ $   &  3277    &0.12 &	0.47 & 0.01  & 0.89  & 4104    & 0.16  & 0.79 & 0.02 & 1.43 \\
$2_5^+ $   &  3860    &0.34 &	1.03 &--	    & 2.05  &             &           &   &          &  \\
$1_2^+ $   &  4322    &0.34 &	--      &  --  & 0.55      &             &           &   &          &  \\
$1_4^+ $   &  4841    &0.21 &	0.01 &  --  & 0.35      &             &           &   &          &  \\\hline
&  \multicolumn{4}{r|}{Total  $\sigma_{\text{theo}}$} &5.87 &  & \multicolumn{3}{r|}{Total  $\sigma_{\text{theo}}$} &2.64\\
  \end{tabular}
\end{ruledtabular}
\end{table*}

\setlength\extrarowheight{2pt}
\begin{table*}[t]
  \caption{Calculated spectroscopic factors and cross sections for the states  populated in the   \ppn~ reaction.  \label{tab:ppn}}
\begin{ruledtabular}
 \begin{tabular}{|c||c|cccc|c||c|cccc|c|}
                                                          & \multicolumn{6}{c||}{\bf{SPDF-MU}}        &  \multicolumn{6}{c|}{\bf{VS-IMSRG}}  \\\hline
\multirow{2}{*}{{\bf J}\bs{$^\pi$}} & \multirow{2}{*}{{\bf E(keV)}} &\multicolumn{4}{c|}{\bs{$C^2S~$}} &{\bs{$\sigma_{\text{\bf theo}}$}}  &   \multirow{2}{*}{{\bf E(keV)}}  &\multicolumn{4}{c|}{\bs{$C^2S~$}} & \bs{$\sigma_{\text{\bf theo}}$} \\
 & &   \bs{$0p_{1/2}$}& \bs{$0p_{3/2}$} & \bs{$0f_{5/2}$} & \bs{$0f_{7/2}$}  &  {\bf (mb)} &   &  \bs{$0p_{1/2}$}& \bs{$0p_{3/2}$} & \bs{$0f_{5/2}$} & \bs{$0f_{7/2}$}   & {\bf (mb)}   \\\hline
$0_1^+$	 & 0	         & 0.57 & --	&  --	       &  --	       &  6.19   & 0           &   0.43  & --  & -- & -- & 4.74\\
$2_1^+$	 & 1291	 & --	     &  0.73 &   0.05 &  --	       &  7.29   & 1328      & --        & 0.83 & --  & -- & 7.95 \\
$0_2^+$	 & 2115	 & 0.28 & --	&  --	       &  --	       &  2.28   & 2317      &  0.38   & -- &  --  & --   & 3.01\\
$1_1^+$	 & 2643	 & --      & 0.91	&  --	       &  --	       &  7.05    & 2864     &  --       & 0.90 & -- & --  & 6.70 \\
$4_1^+$	 & 2651	 & --      &  --	&   --        &  0.93     &  5.33   & 3201    &  --    & --  & -- & 0.96  & 5.54\\
$2_2^+$	 & 2676	 & --	    &  0.25 &   0.05  &  --	       &  2.34   &          &           &          &        &       &   \\
$2_3^+$	 & 2986	 & --	    &  0.73 &   0.02  &  --	       &  5.47   & 3605   & -- & 0.63 & -- & -- & 4.09 \\
$3_2^+$	 & 3631	 & --	   &  --	&   0.05  &  0.40     &  2.34   &          &           &          &        &       &   \\
$4_2^+$	 & 3644	 & --	   &  --	&   --      &  0.11     &  0.56   &          &           &          &        &       &   \\
$3_3^+$	 & 3698	 & --	   &  --	&   0.03  &  0.70    &   3.79   & 4428 & -- & -- & -- & 1.05 & 9.44\\
$4_4^+$	 & 4481	 & --	   &  --	&   --      &  0.23    &   1.26   &          &           &          &        &       &   \\
$1_3^+$	 &         	 &	   & 	       &             &        &                & 4819  &   0.15  & --      &   --     & --      & 0.67  \\
$1_5^+$	 & 4983	 & 0.01  &  0.14&   --      &  --        &  0.70   &          &           &          &        &       &   \\\hline
  &   \multicolumn{5}{r|}{Total  $\sigma_{\text{theo}}$} &44.43  &   \multicolumn{5}{r|}{Total  $\sigma_{\text{theo}}$} & 42.14 \\
 \end{tabular}
\end{ruledtabular}
\end{table*}

To get an insight on the spin and parity of the  observed levels,   
%the predicted population of low-lying states in \ts{50}Ar following proton- and neutron-knockout was investigated.  For this, 
single-particle theoretical cross sections  were computed in  the  DWIA framework~\cite{Wakasa_PPNP_96_2017} using the Bohr-Mottelson single-particle potential~\cite{Bohr-Mottelson}. For the optical potentials of the distorted waves, the microscopic folding model~\cite{Toyokawa_PRC_88_2013} with the Melbourne G-matrix interaction~\cite{Amos_ANP_25_2000} and  calculated nuclear density was employed.  The  Franey-Love effective proton-proton interaction was adopted~\cite{Franey_PRC_31_1985} and the spin-orbit part of each distorting potential was disregarded.
Cross sections at different beam energies were calculated to take into account the energy loss of the beam in the thick target. 
The calculated single-particle cross sections  were multiplied by the spectroscopic factors calculated for the reactions in each theoretical framework. 
%\begin{figure}[t]
% \includegraphics[angle=-90,width=0.48\textwidth]{SigmaPlotV5.eps}%
% \caption{Comparison between the energy levels and cross sections  a) experimentally observed, and the b), c) theoretical calculations for the \ptp~ reaction. The calculated contribution from each orbital is depicted. \label{fig:sigmaplot_p2p}}
%\end{figure}
Tables~\ref{tab:ptp} and \ref{tab:ppn} show the obtained results  for the \ptp~ and \ppn~ reactions, respectively. 

The calculated  ground state cross section for both reactions is much lower than the experimental values. As already mentioned this is due to  the non-observation of states decaying directly to the $0^+_1$ state, which results in an over-estimation of the  experimental cross section. 
For the case of the \ptp~ reaction, the  SDPF-MU and VS-IMSRG calculations  predict a  cross section to the \stwop  state of  1.0~mb and 0.62~mb, respectively, in reasonable agreement with the experimental value of 0.8(2)~mb.
At higher energies the SPDF-MU calculation suggest the population of the $2^+_3$,  $2^+_4$,  and  $2^+_5$ states. 
Although high cross sections are also predicted for the  $1^+_2$ and $1^+_4$ states, they would decay preferentially to the ground-state, therefore its correspondence to any experimental level is unlikely.  
They may, however, account  for the seeming too high experimental population of the ground state when compared to calculated cross sections.
The VS-IMSRG calculation, on the other hand, only indicates the population of the $2^+_4$ and $4^+_1$ states.  
The fact that the VS-IMSRG calculations only predicts two states  with sizable $sd$-proton cross-sections is  related with the reduced model space, which prevents proton $pf- sd$ excitations.
This in turn, highlights the importance of such excitations in the population of low-lying states. They will be investigated in the future with a newly developed cross-shell VS-IMSRG approach~\cite{Miyagi_arxiv_2020}.
In spite of the differences between the models, they both point out  that the  \ptp~ reaction mostly populates  $2^+$ states. The   experimental levels at 2744(20)~keV, 3053(23)~keV, and 3771(34)~keV, observed in this reaction, are in fair agreement with the  predictions for the $2^+_3$,  $2^+_4$,  and  $2^+_5$  states in the SDPF-MU model. We therefore tentatively assign this spin and parity to these states. 
The  level at 2744(20)~keV has been previously suggested to be the \sfourp~\cite{Steppenbeck_PRL_114_2015}. Although the  SDPF-MU calculations favors a  $2^+$ assignment, the comparison with the VS-IMSRG results make it also compatible with the \sfourp. 
Furthermore,  the  population of this state in the \ppp~ reaction favors a \sfourp assignment. Therefore  a  $(2^+,4^+)$ assignment is left open for this state. 
It is worth mentioning  that the state at 1976(15)~keV has a negligible cross section for the \ptp~ reaction, which is consistent with the theoretical predictions for the $0^+_2$ state.  
The agreement between the SDPF-MU and VS-IMSRG calculations on the energy and spectroscopic factor of the $0^+_2$ state suggest that it is spherical in nature, as the VS-IMSRG does not properly account for deformed low-lying states~\cite{Cortes_PLB_800_2019,Taniuchi_Nature_569_2019}.

%\begin{figure}[t]
% \includegraphics[angle=-90,width=0.48\textwidth]{SigmaPlot_ppnV5.eps}%
% \caption{Same as Fig.~\ref{fig:sigmaplot_p2p} for the case of the \ppn~ reaction.\label{fig:sigmaplot_ppn}}
%\end{figure}

For the \ppn~reaction the  theoretical models predict a \stwop state cross section of  $\sim7 - 8$~mb, while the experimental value is 15(4)~mb.  In this case, the experimental over-estimation comes from the impossibility to distinguish between the direct and indirect reactions in this channel. 
The next states with the  higher predicted cross section  are the $1^+_1$, \sfourp, and $2^+_3$ states in both calculations.  In the VS-IMSRG,  population to the $3^+_3$ state at 4428~keV is also predicted.  As previously noted, the  $1^+_1$  state would most probably decay directly to the ground-state. Furthermore, the $3^+_3$ is not predicted by the VS-IMSRG to be populated in the \ptp~reaction, so it is  improbable that it corresponds to an experimental level.  The ambiguity between the $2^+$ and $4^+$ characters for the state at 2744(20)~keV observed in this reaction is therefore maintained.

\begin{figure}[t]
 \includegraphics[angle=-90,width=0.45\textwidth]{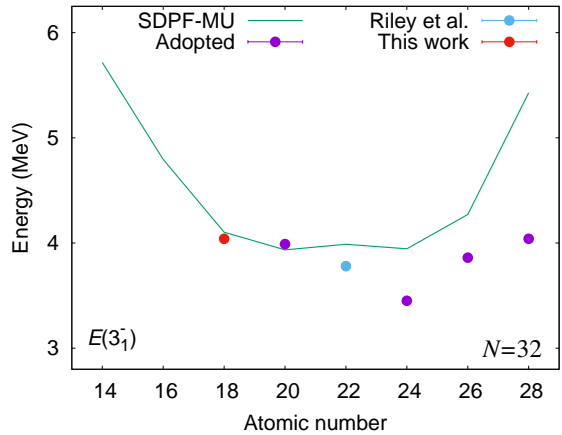}%
 \caption{Systematics of $E(3^-)$ for even-even $N=32$ isotones. The circles represent the available data~\cite{NNDC,Riley_PRC_96_2017}, including the value for \ts{50}Ar reported in this work. The solid line shows the SDPF-MU calculations. 
   \label{fig:systematics}}
\end{figure}

Finally,  theoretical predictions of the  systematic of $3^-_1$ states for the $N=32$ isotones have been obtained using the SDPF-MU calculations and confronted to available data~\cite{NNDC,Riley_PRC_96_2017} including  the state at 4040(44)~keV  obtained in this work,  as shown in  Fig.~\ref{fig:systematics}.
The $E(3^-_1)$ for \ts{50}Ar is comparable in magnitude to the one of \ts{52}Ca,  
%, and is the highest reported so far along the $N=32$ isotones. 
and the theoretical predictions show a good agreement with both  isotopes, reinforcing the spin and parity assignments.
However, the calculations  overestimate the  $E(3^-_1)$ as $Z$ increases. %This is particularly clear for the \sthreem state of \ts{60}Ni.
For nuclei around Ca, the Fermi surface is located near the $Z=20$ shell gap,  therefore proton excitations require  less energy. This is reflected in the calculations by the low \sthreem levels predicted  for \ts{50}Ar, \ts{52}Ca, and \ts{54}Ti, where the calculations show a good agreement with the data. 
Going towards the Si isotopes, the  excitation from the $p$ shell to the $sd$ shell becomes likely. 
This possible  excitation is not taken into account in the calculations, which in turn increases the predicted $E(3^-_1)$ energies.
On the other side,  towards higher $Z$, the experimental levels are rather stable around 4~MeV, but the calculations are  not able to reproduce them.
%In this case the most probably cause are the neutron excitations. 
In Ni, proton excitations  from the $sd$ to the $pf$ shells, as well as neutron excitations  from the $pf$ to the $sdg$ shells may  contribute. In particular, it has been reported that  the neutron excitations from the $pf$ to the $sdg$ shell are not well reproduced due to a too large shell gap between $pf$ and  $sdg$ shells, and it has been suggested that it is necessary to lower the $sdg$ shell by 1~MeV to reproduce the negative parity states of Ni isotopes~\cite{Shimizu_PLB_753_2016,Utsuno_PTPS_196_2012}.

%\appendix
\section{Summary}

Low-lying levels of \ts{50}Ar have been  investigated by proton- and neutron-knockout reactions,  inelastic proton scattering, and multinucleon removal reactions.
Based on the $\gamma\gamma$ analysis, a level scheme was constructed, including five newly  observed transitions  among which a candidate for  the \sthreem state has been reported for the first time. 
The experimental level scheme was compared to theoretical calculations performed in the SDPF-MU shell model, as well as the ab initio VS-IMSRG approach. Both calculations predict similar level schemes for \ts{50}Ar. 
%Calculated spectroscopic factors were combined with single-particle  cross sections calculated in the DWIA framework to obtain theoretical cross sections. 
Theoretical cross sections for the  \ptp~ and \ppn~ were compared to the  observed ones, to infer the spin and parity of the states. 
Two of the  newly observed states were tentatively assigned a $(2^+)$ spin and parity, and it was shown that the  state previously suggested to be the \sfourp  could  also correspond to a $2^+$ state. 

% is suggested  as a candidate for the  $(0^+_2)$ state. 

Overall, both theoretical calculations provide consistent results and a relatively good agreement with the experimental data for both the \ptp~ and \ppn~ reactions.
This   emphasizes the sub-shell closure at $N=32$ and our understanding of shell evolution in this region. 
The remaining differences among calculations most likely arise from the reduced proton and neutron spaces employed in the VS-IMSRG and highight their importance in the understanding of the low-lying structure of \ts{50}Ar.


\begin{acknowledgments}
  We thank the RIKEN Nishina Center accelerator staff and the BigRIPS team for the stable operation of the high-intensity  Zn beam and for the preparation of the secondary beam  setting. 
  We thank S.~R.~Stroberg for very useful discussions.
  
  This work has been supported by the Grant-in-Aid for Scientific Research  JP16K05352, the RIKEN Special Postdoctoral Researcher Program, Colciencias - convocatoria 617 becas doctorados nacionales, the Ministry of Science and Technology of Vietnam through the Physics Development Program Grant No.{\DJ}T{\DJ}LCN.25/18,  HIC for FAIR, the Croatian Science Foundation under projects no. 1257 and 7194, the  GINOP-2.3.3-15-2016-00034 and  K128947 projects, the NKFIH (128072), the Spanish Ministerio de Econom\'ia y Competitividad under Contract No. FPA2017-84756-C4-2-P,  the NRF grants No. 2018R1A5A1025563 and 2019M7A1A1033186 funded by the Korean government,  the JSPS KAKENHI Grant No. 18K03639, MEXT as ``Priority issue on post-K computer'' (Elucidation of the fundamental laws and evolution of the universe), the Joint Institute for Computational Fundamental Science (JICFuS), the Ram\'on y Cajal program RYC-2017-22781 of the Spanish Ministry of Science, Innovation and Universities, the Natural Sciences and Engineering Research Council (NSERC) of Canada, the Deutsche Forschungsgemeinschaft (DFG, German Research Foundation) -- Project-ID 279384907 -- SFB 1245 and grant No. BL 1513/1-1, the PRISMA Cluster of Excellence, and the BMBF under Contracts No. 05P15RDFN1, 05P18RDFN1 and 05P19RDFN1. TRIUMF receives funding via a contribution through the National Research Council Canada. Computations were performed with an allocation of computing resources on Cedar at WestGrid and Compute Canada, and on the Oak Cluster at TRIUMF managed by the University of British Columbia, Department of Advanced Research Computing (ARC).
 The development of MINOS was supported by the European Research Council through the ERC Grant No. MINOS-258567.

\end{acknowledgments}

%\bibliographystyle{CUEDbiblio-url2}

%\bibliography{References}

\end{document}